\documentclass[twocolumn,english,journal]{IEEEtran}
\usepackage{listings}
\usepackage{babel}
\usepackage{float}
\usepackage{multirow}
\usepackage{amsmath}
\usepackage{graphicx}

\makeatletter

\usepackage{stfloats}

\usepackage[capitalise]{cleveref}
\usepackage{algorithm}
\usepackage{algpseudocode}
\newtheorem{mytheorem}{Theorem}
\newtheorem{myremark}{Remark}

\ifCLASSOPTIONcompsoc
\usepackage[caption=false,font=normalsize,labelfont=sf,textfont=sf]{subfig}
\else
\usepackage[caption=false,font=footnotesize]{subfig}
\fi

\addto\captionsenglish{}

\addto\captionsenglish{}

\Crefname{algorithm}{\textbf{Algorithme}}{\textbf{Algorithmes}}

\makeatother

\begin{document}

\title{Utility Max-Min Fair Link Adaptation in IEEE 802.11ac Downlink Multi-User}

\author{Ali A. Khavasi, Mojtaba Aajami, Hae-Ryeon Park and Jung-Bong Suk%
\thanks{Manuscript received January 31, 2014 ----------------.%
}%
\thanks{Ali A. Khavasi\textsuperscript{1}, Mojtaba Aajami\textsuperscript{2},
Hae-Ryeon Park\textsuperscript{3} and Jung-Bong Suk\textsuperscript{4}
are with the Department of Computer and Telecommunications Engineering,
Yonsei University, Korea (e-mail: \{khavasi\textsuperscript{1},maajami\textsuperscript{2},jbsuk\textsuperscript{4}\}@yonsei.ac.kr,
haerin20\textsuperscript{3}@hanmail.net).%
}%
\thanks{%
}}
\maketitle
\begin{abstract}
In this letter, we propose a novel model and corresponding algorithms
to address the optimal utility max-min fair link adaptation in Downlink
Multi-User (DL-MU) feature of the emerging IEEE 802.11ac WLAN standard.
Herein, we first propose a simple yet accurate model to formulate
the max-min fair link adaptation problem. Furthermore, this model
guarantees the minimum utility gain of each receiver according to
its requirements. In the second step, we show that the optimal solution
of the proposed model can be obtained in polynomial time, and then
the solution algorithms are proposed and analyzed. The simulation
results demonstrate the significant achievement of the proposed utility-aware
link adaptation approach in terms of max-min fairness and utility
gain compared to utility-oblivious schemes.\end{abstract}

\section{Introduction}

\IEEEPARstart{U}{tility} max-min (UMM) fairness has been applied
in numerous applications since the initial work by Cao and Zegura
\cite{cao1999util-maxmin}, who attempt to find an optimal resource
allocation such that the utilities of all receivers are equal. Hence,
the social welfare in the network is equalized. More precisely, this
can be achieved by solving the following optimization problem:

\begin{equation}
\mathbf{\mathbf{UMM}:}\max_{\rho_{r}^{i}}\min_{r\in R}U_{r}(\rho_{r}^{i})\label{eq:UMM-Optimization}
\end{equation}

\begin{equation}
s.t.\begin{cases}
\sum_{r=1}^{R}p_{r}^{i}\leq P_{T}\\
\forall m_{r}^{j},\, j=1\ldots M
\end{cases}
\end{equation}
where $r\,(r=1,2,..R|$$R:$ the number of receivers$)$ is the receiver
index, $\rho_{r}^{i}\in\left\{ \rho_{r}^{1},\rho_{r}^{2},\ldots\right\} $
is defined as the $i\textrm{th}$ policy ($i=1,2,\ldots$) of all
available link adaptation policies, which further determines the selected
Modulation and Coding Scheme (MCS) $m_{r}^{j}$ ($j=1\ldots M$$\textrm{\ensuremath{|M:}}$
the number of total available MCS) and the allocated power $p_{r}^{i}$
($p_{r}^{i}\in[0..P_{T}]\textrm{\ensuremath{|P_{T}:}the total available transmission power})$,
and finally $U_{r}(\rho_{r}^{i})$ is the utility of receiver $r$
due to the applied policy $\rho_{r}^{i}$. Through this letter, we
assume each $U_{r}$ is a positive, non-decreasing and normalized
utility function of $\rho_{r}^{i}$ \cite{WH.Wang-app-alg-fair}.
The max-min problem $\mathbf{\mathbf{UMM}}$, has a unique optimal
solution under a critical assumption of strict high Signal to Interference
and Noise Ratio (SINR) and concavity of the receiver utilities $U_{r}(\rho_{r}^{i})$.
However, there are three impediments in applying the problem formulation
(\ref{eq:UMM-Optimization}). First, the concavity assumption does
not hold for real-time applications such as video transmission and
Voice over IP (VoIP). The utility of these applications can be approximated
by a step or a sigmoid function. Moreover, the high SINR assumption
is not guaranteed in WLANs. As a result, the optimization problem
(\ref{eq:UMM-Optimization}), in general, is a non-convex problem
and cannot be reduced to Geometric Programming (GP). Indeed, it belongs
to the class of problems known as Complementary GP that is, in general,
an NP-hard or that may require a very complex algorithm to solve \cite{Chiang-Non-Convex}.
Second, real-time applications that need a certain amount of transmission
resources to provide the minimum required satisfaction (acceptance)
suffer from so called ``bandwidth starvation''. Third, in applying
the theoretical formulas to predict the bit rate and corresponding
Frame Error Rate (FER) associated with the link adaptation policy
$\rho_{r}^{i}$, there are simply too many ways in which the observed
measurements and the actual performance fail to match the predictions
of the theory \cite{halperin2010predictable}. Through this letter,
it is shown that the proposed scheme efficiently addresses these three
impediments.

The IEEE 802.11ac is developed by extending the air\textendash{}interface
techniques of the 802.11n standard. Among the offered features, the
DL-MU exploits the beamforming technique to steer signal maxima on
certain receivers while suppressing, or at least greatly reducing
the corresponding interference over the others. The channel state
information (CSI) of the target receivers needs to be known at the
transmitter, and it is provided by the channel estimation module in
the standard \cite{IEEE802.11acDraft-6627922}. 

In this letter, we propose a novel, practical, and optimal link adaptation
algorithm in order to achieve the max-min fair utility distribution
among the receivers in the IEEE 802.11ac DL-MU.

\section{Proposed Model and Theoretical Explanations}

We first start with devising a model to predict the FER corresponding
to each selected policy $\rho_{r}^{i}$ in order to address the third
aforementioned impediment. In \cite{halperin2010predictable}, the
authors propose a model, which, by using the concept of effective
SNR, performs FER prediction for each choice of $\rho_{r}^{i}$. We
further extend their model to include beamforming and to match the
802.11ac specification \cite{IEEE802.11acDraft-6627922}, as follows.
In accordance with the precoding scheme applied by the transmitter,
the SNR of $l$th subcarrier of the receiver $r$ (denoted by $\varsigma_{l_{r}}$),
is obtained based on the available CSI at the transmitter. For instance,
in the case of zero-forcing, the average SNR value of the certain
subcarrier $l$ is calculated as

\begin{equation}
\varsigma_{l_{r}}=\frac{p_{r}^{i}\times||\mathbf{h}_{l_{r}}\mathbf{w}_{l_{r}}||^{2}}{\sigma^{2}}\label{eq:SNR}
\end{equation}
where $||\cdot||$ is the norm of the vector, $\sigma^{2}$ is the
observed noise variance at the receiver, and $\mathbf{h}_{l_{k}}=(h_{l_{r1}},h_{l_{r2}},\cdots,h_{l_{rR}})$
is the vector of channel coefficients from $R$ antennae of the transmitter
to the single antenna of receiver $r$. Additionally, $\mathbf{w}_{l_{r}}$
is the unit-norm beamforming weight vector ($\mathbf{w}_{l_{r}}=(w_{l_{r1}},w_{l_{r2}},\cdots,w_{l_{rR}})$)
for receiver $r$. Using (\ref{eq:SNR}), the BER of the wideband
downlink channel $b^{m_{r}^{j}}$ for modulation of $m_{r}^{j}$ is
estimated by

\begin{equation}
b^{m_{r}^{j}}=\frac{1}{L}\sum_{l=1}^{L}BER^{m_{r}^{j}}(\varsigma_{l_{r}})\label{eq:BER}
\end{equation}
where $L$ is the total number of data subcarriers, and the BER of
the subcarrier $l$ is denoted by $BER^{m_{r}^{j}}(\varsigma_{l_{r}})$
as a function of the symbol SNR (here, $\varsigma_{l_{r}}$) in the
narrowband channel \cite{Goldsmith05wirelesscommunications}. Note
that $b^{m_{r}^{j}}$ is the uncoded bit error probability for the
modulation \textbf{$m_{r}^{j}$}. In order to obtain the upper bound
of the coded FER, we use the union bound on the first-event error
probability $E_{u}^{c}$. By assuming that the frame is transmitted
using the convolutional code $c$, then $E_{u}^{c}=\sum_{d=d_{free}}^{\infty}a_{d}.E_{d}$,
where $d_{free}$ is the free distance of the code $c$, $a_{d}$
is the total number of error events of weight $d$, and $E_{d}$ is
the probability of an incorrect path of the hamming distance $d$
when it diverges from the correct path and then re-merges with it
sometime later. The value of $a_{d}$ for a specific convolutional
code is generated using its transfer function, and $E_{d}$ is obtained
by

\begin{equation}
E_{d}=\begin{cases}
\sum_{k=\frac{(d+1)}{2}}^{d}\binom{d}{k}\cdot(b^{m_{r}^{j}})^{k}\cdot(1-b^{m_{r}^{j}})^{d-k},\:\textrm{if d is odd} & \textrm{}\\
\\
\frac{1}{2}\cdot\binom{d}{\frac{d}{2}}\cdot(b^{m_{r}^{j}})^{\frac{d}{2}}\cdot(1-b^{m_{r}^{j}})^{\frac{d}{2}}+\\
\,\sum_{k=\frac{(d+1)}{2}}^{d}\binom{d}{k}\cdot(b^{m_{r}^{j}})^{k}\cdot(1-b^{m_{r}^{j}})^{d-k},\:\textrm{if d is even} & \textrm{}
\end{cases}
\end{equation}
Therefore, by applying the policy $\rho_{r}^{i}$, the predicted upper
bound of FER on wideband channel (here, $FER_{eff}$) for a frame
of length $L_{f}$ is

\begin{equation}
FER_{eff}\leq1-(1-E_{u}^{c})^{L_{f}}\label{eq:upperboundofFER}
\end{equation}

Using (\ref{eq:BER}) and (\ref{eq:upperboundofFER}), we employ a
method similar to \cite{halperin2010predictable} in order to construct
the FER prediction model. In this method, based on the available CSI,
for each non-overlapping range of the transmit power level $p_{r}^{\Xi}$
($\Xi\in(p^{min},p^{max}]$$|min,max\in[1,2,...]$ and $min<max$),
there exists an MCS \textbf{$m_{r}^{\Theta}$} ($\Theta\in[1\ldots M]$)
such that we predict $\leq FER_{eff,r}^{\Xi}$ frame error rate. Note
that for each $p_{r}^{\Xi}$ (and therefore corresponding effective
SNR) only one MCS \textbf{$m_{r}^{\Theta}$} results in FER between
$1-(\frac{1}{2})^{L_{f}}$ (useless) and $0$ (lossless). We assume
that $p_{r}^{\Xi}$ has the maximum value in the range, unless stated
otherwise. Subsequently, the utility value $U_{r}(\rho_{r}^{\Xi})$
associated with each tuple $\rho_{r}^{\Xi}(p_{r}^{\Xi},m_{r}^{\Theta},FER_{eff,r}^{\Xi})$
is calculated using its specific function. On-site calibration is
necessary to obtain tuples of the table $\Phi_{r}(p_{r}^{\Xi},m_{r}^{\Theta},FER_{eff,r}^{\Xi})$
of length $L_{\Phi}$ (the number of tuples) for each receiver $r$. 

Now, we propose a new model to address the second aforementioned impediment.
First, we attempt to determine the policies that satisfy the user-defined
constraints. Such constraints are crucial, especially for real-time
applications that need a certain amount of transmission resources
to provide the required quantity of utility. We are interested in
the policy $\rho_{r}^{min}$ with the allocated power $p_{r}^{min}$
for each receiver $r$ such that the minimum utility requirement is
satisfied. Accordingly, the minimization problem with the object function
of $U_{r}(\rho_{r}^{\Xi})$ can be expressed as

\begin{equation}
\rho_{r\in R}^{min}=\min_{p_{r}^{\Xi}}(\arg\min_{\rho_{r}^{\Xi}\in\Phi_{r}}\quad U_{r}(\rho_{r}^{\Xi}))\label{eq:minUtil}
\end{equation}

\begin{equation}
s.t.\begin{cases}
\sum_{all\, receivers}p_{r}^{\Xi}\leq P_{T}\\
p_{r}^{\Xi}\geq0\\
\forall m_{r}^{\Theta},\,\Theta=1\ldots M\\
U_{min,r}\leq U_{r}(\rho_{r}^{\Xi})\leq1
\end{cases}\label{eq:minUtil-constraints}
\end{equation}
where $U_{min,r}$ is the user-defined minimum value of the required
utility. Once the minimum required utility is granted to each receiver
by (\ref{eq:minUtil}), we attempt to maximize all utilities equally
using the $\mathbf{\mathbf{UMM}}$ problem definition. Therefore,
we define the utility max-min fair link adaptation problem with the
object function of $O_{r}^{\Xi}(U_{r}(\rho_{r}^{\Xi}))=(U_{r}(\rho_{r}^{\Xi})-U_{min,r})$
as

\begin{equation}
\max_{\rho_{r}^{\Xi}\in\Phi_{r}}\min_{r\in R}(U_{r}(\rho_{r}^{\Xi})-U_{min,r})\label{eq:utility-max-min-model}
\end{equation}

\begin{equation}
s.t.\begin{cases}
\sum_{all\, receivers}p_{r}^{\Xi}\leq P_{T}\\
p_{r}^{\Xi}\geq p_{r}^{min}\\
\forall m_{r}^{\Theta},\,\Theta=1\ldots M\\
U_{min,r}\leq U_{r}(\rho_{r}^{\Xi})\leq1
\end{cases}
\end{equation}
We state the polynomial time complexity of the optimal solution to
the problem (\ref{eq:utility-max-min-model}) in the following theorem.

\begin{mytheorem}The optimal solution of the addressed problem (\ref{eq:utility-max-min-model})
can be obtained in $O(R\cdot L_{\Phi}\cdot\log(L_{\Phi}))$.\end{mytheorem}
\begin{IEEEproof}
If the resources can be allocated in advance, the optimal max-min
fair solution is achieved by using the progressive filling algorithm
\cite{bertsekas1992data}. In this case, the algorithm starts with
all utility gains equal to 0 and increases them together at the same
pace, until one or several limits are reached. It continues to increase
the gains for other receivers until it is not possible to further
increase any utility. If the policies are sorted descending by their
utility values (gains), in the worst case it will take $O(R\cdot L_{\Phi})$
for the algorithm to accomplish this task. If this is not the case,
policies can be sorted in advance, and therefore the time complexity
is on the order of $O(R\cdot L_{\Phi}\cdot\log(L_{\Phi}))$.

It is worth noting that the multi-rate WLAN suffers from performance
anomaly when low-rate streams occupy most of the shared channel time.
According to \cite{cantieni2005performance}, allocating same channel
occupation time to the different streams solves this issue and maintains
channel time fairness between the streams. In the 802.11ac DL-MU transmission,
all streams have the same TXOP time which implies the channel time
fairness amongst them in DL-MU transmissions.
\end{IEEEproof}

\section{Solution Algorithms}

\begin{algorithm}[t]
\caption{Finding the policy $\rho_{r}^{min}$ \label{alg:MinUtil}}

\begin{lstlisting}[basicstyle={\footnotesize},language=Matlab,mathescape=true,tabsize=2]
 1	for r=1:R
 2		$\Phi_{r}^{sel}$ = select array of tuples from $\Phi_{r}$ 
 								where: $\Phi_{r}^{\Xi}.U_{r}^{\Xi}\geq U_{min,r}$;
 3		$\rho_{r}^{min}$ = min($\Phi_{r}^{sel}.p_{r}^{\Xi})$;
 4		remove tuples with $U_{r}^{\Xi}\leq \rho_{r}^{min}.U_{r}^{\Xi}$ from $\Phi_{r}^{sel}$;
 5	end
 6	if $\overset{R}{\underset{r=1}{\sum}}\rho_{r}^{min}.p_{r}^{\Xi}\leq P_{T}$ 
 7		return $\Phi_{\{r\in R\}}^{sel}$ and $\rho_{\{r\in R\}}^{min}$;
 8	end
\end{lstlisting}
\end{algorithm}
\begin{algorithm}[t]
\caption{Utility max-min fair solution to the problem (\ref{eq:utility-max-min-model})\label{alg:Max-minUtil}}

\begin{lstlisting}[basicstyle={\footnotesize},language=Matlab,mathescape=true,tabsize=2]
 1		$\textrm{index}_{\Phi}$(1:R)=2; $\rho_{\{r\in R\}}^{sel}=\rho_{\{r\in R\}}^{min}$; $O_{\{r\in R\}}$=Inf(R,1);
 2		for i=1:R
 3			if $\Phi_{i}^{sel}.L_{\Phi}\geq 2$
 4				$O_{i}$ = $\Phi_{i}^{sel}\{\textrm{index}_{\Phi}(i)\}$.$U_{i}^{\Xi}$ - $\rho_{i}^{min}$.$U_{i}^{\Xi}$;
 5			end
 6		end
 7		while true
 8			[newmin, j] = min($O_{\{r\in R\}}$);
 9			if newmin == Inf
10				return $\rho_{\{r\in R\}}^{sel}$;
11			end
12			if $(\Phi_{j}^{sel}\{\textrm{index}_{\Phi}(j)\}.p_{j}^{\Xi}+\underset{r\neq j}{\sum}\rho_{r}^{sel}.p_{r}^{\Xi})\leq P_{T}$
13				$\rho_{j}^{sel}$ = $\Phi_{j}^{sel}\{\textrm{index}_{\Phi}(j)\}$;
14				$\textrm{index}_{\Phi}(j)$ = $\textrm{index}_{\Phi}(j) + 1$;
15				if $\textrm{index}_{\Phi}(j)\leq \Phi_{j}^{sel}.L_{\Phi}$
16					$O_{j}$ = $\Phi_{j}^{sel}\{\textrm{index}_{\Phi}(j)\}.U_{j}^{\Xi}-\rho_{j}^{min}.U_{j}^{\Xi}$;
17				else
18					$O_{j}$ = Inf;
19				end
20			else
21				$O_{j}$ = Inf;
22			end
23		end
\end{lstlisting}
\end{algorithm}

Theorem 1 states the existence of the polynomial-time solution algorithms
to problem (\ref{eq:utility-max-min-model}) and, therefore, the first
aforementioned impediment is addressed by the proposed scheme. Here,
we propose two algorithms to obtain the solutions to the problems
(\ref{eq:minUtil}) and (\ref{eq:utility-max-min-model}). First,
we propose \textbf{Algorithm \ref{alg:MinUtil},} in MATLAB pseudo-code,
which provides the solution to the problem (\ref{eq:minUtil}). Then,
\textbf{Algorithm \ref{alg:Max-minUtil}} gives the solution to the
problem (\ref{eq:utility-max-min-model}) by exploiting the previously
obtained result. In the following theorem, we state that the given
solution is optimal.

\begin{mytheorem}\textbf{Algorithm \ref{alg:Max-minUtil}} provides
the optimal solution to the problem (\ref{eq:utility-max-min-model})
in $O(R\cdot L_{\Phi}\cdot\log(L_{\Phi}))$.\end{mytheorem}
\begin{IEEEproof}
The progressive filling algorithm guarantees a max-min fair distribution
of the utilities. We show that \textbf{Algorithm \ref{alg:Max-minUtil}}
follows the problem formulation (\ref{eq:utility-max-min-model})
and the progressive filling algorithm. The constraint $\sum_{\textrm{all receivers}}p_{r}^{\Xi}\leq P_{T}$
is ensured in line 12. Note that the other constraints are already
met in tables $\Phi_{r}$ and $\Phi_{r}^{sel}$, which is previously
defined in line 2 of \textbf{Algorithm \ref{alg:MinUtil}}. Line 1
initiates the progressive filling by setting all utility gains to
0 ($\rho_{r}^{sel}=\rho_{r}^{min}\Longrightarrow O_{r}^{sel}(U_{r}(\rho_{r}^{sel})))$$=U_{r}(\rho_{r}^{sel})-U_{min,r}=0$).
Lines 7-23 aim to maximize the objective function $O_{r}^{\Xi}(U_{r}(\rho_{r}^{\Xi}))$
by increasing all gains iteratively at the same pace. Lines 18 and
21 ensure the exclusion of the receivers that reach the capacity limits.
The iteration (lines 7-23) continues until it is not possible to further
increase the objective function (lines 9-11). 
\end{IEEEproof}
\begin{myremark} \textbf{Algorithm \ref{alg:MinUtil}} is on the
order of $O(R\cdot L_{\Phi})$ and \textbf{Algorithm \ref{alg:Max-minUtil}}
runs in $O(R\cdot L_{\Phi}\cdot\log(L_{\Phi}))$. \end{myremark}

\section{Simulation Results}

\begin{figure}[t]
\includegraphics[width=1\columnwidth]{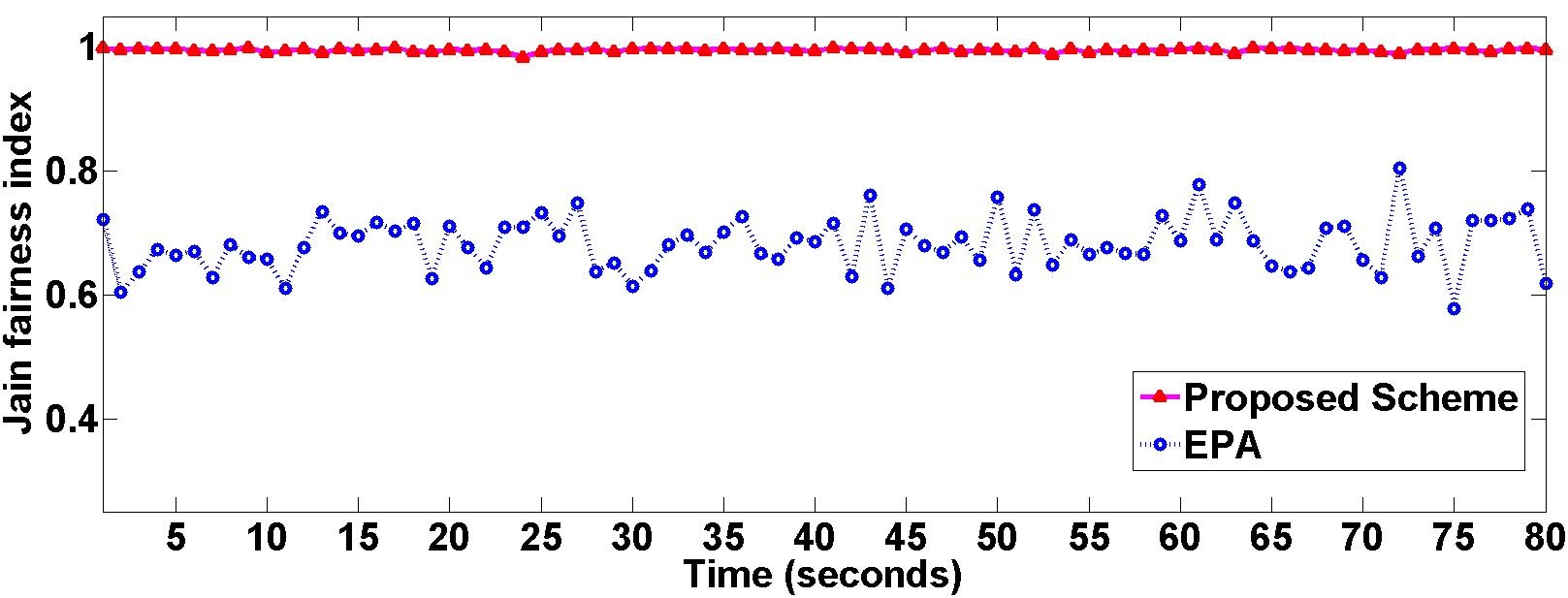}

\caption{Jain's fairness index for two compared schemes: Proposed and EPA\label{fig:Jain's-Fairness-Index}}
\end{figure}

\begin{figure}[t]
\includegraphics[width=1\columnwidth]{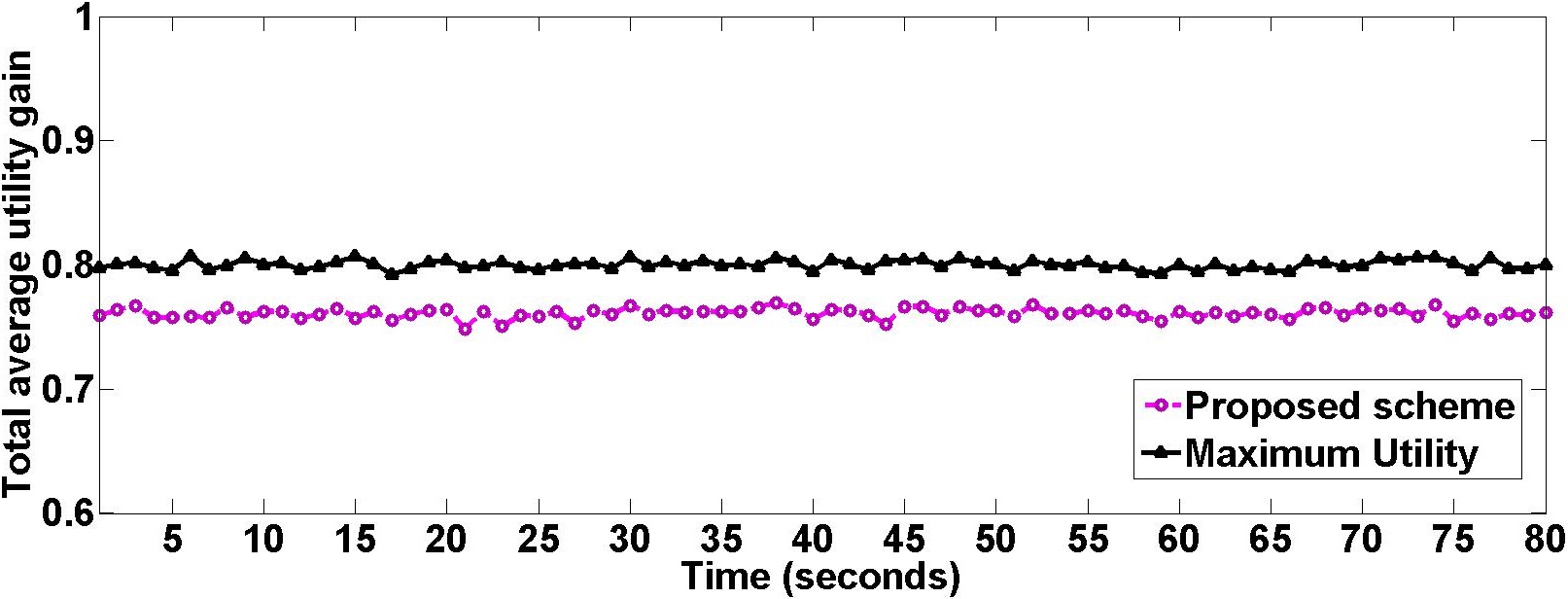}

\caption{Utility average of two compared schemes: Proposed and Maximum Utility\label{fig:Total-average-of} }
\end{figure}
\begin{table*}[t]
\caption{Utility Functions of Different Applications\label{tab:Util-Functions}}

\begin{tabular}{|c|c|c|}
\hline 
{\scriptsize{Application}} & {\scriptsize{Utility Function}} & {\scriptsize{Calibration Parameters}}\tabularnewline
\hline 
\hline 
\multirow{2}{*}{{\scriptsize{VoIP}}} & \multirow{2}{*}{{\scriptsize{$U_{1}^{\Xi}=(1-FER_{eff,r}^{\Xi})\cdot\underset{i=0}{\overset{L_{T}}{\sum}}\alpha_{i}\cdot X(rate(m_{r}^{\theta}),l_{i})$}}} & {\scriptsize{$X(x,l_{i})=\begin{cases}
1 & x\in l_{i}\\
0 & x\notin l_{i}
\end{cases}$, where $l_{i}:\textrm{ required rate of the quality/utility level}$}}\tabularnewline
 &  & {\scriptsize{$L_{T}:\textrm{total number of levels}$, $\alpha_{i}:\textrm{scaling parameter of }l_{i}$}}\tabularnewline
\hline 
\multirow{2}{*}{{\scriptsize{Video Streaming}}} & \multirow{2}{*}{{\scriptsize{$U_{2}^{\Xi}=(1-FER_{eff,r}^{\Xi})\cdot\frac{1}{1+(\frac{1}{\epsilon}-1)\cdot e^{-\beta\cdot rate(m_{r}^{\theta})}}$}}} & {\scriptsize{$\beta=\frac{2\cdot\ln(\frac{1}{\epsilon}-1)}{RATE_{max}}$,
where $RATE_{max}:\textrm{maximal rate requirement}$}}\tabularnewline
 &  & {\scriptsize{$\epsilon:\textrm{minimum of utility gain achievable by }RATE_{min}$,
$RATE_{min}=\textrm{minimal rate requirement}$}}\tabularnewline
\hline 
{\scriptsize{File Transfer}} & {\scriptsize{$U_{3}^{\Xi}=(1-FER_{eff,r}^{\Xi})\cdot\frac{\log(rate(m_{r}^{\theta})+1)}{\log(RATE_{max}+1)}$}} & {\scriptsize{$RATE_{max}:\textrm{maximal rate requirement}$}}\tabularnewline
\hline 
\multirow{3}{*}{{\scriptsize{Online Gaming}}} & \multirow{3}{*}{{\scriptsize{$U_{4}^{\Xi}=(1-FER_{eff,r}^{\Xi})\cdot\frac{1}{1+(\frac{1}{\epsilon}-1)\cdot e^{-\gamma\cdot rate(m_{r}^{\theta})}}$}}} & {\scriptsize{$\gamma=\frac{1}{\underset{i=1}{\overset{N}{\sum}}\frac{t_{i}}{\gamma_{i}}}$,
$\gamma_{i}=\frac{2\cdot\ln(\frac{1}{\epsilon}-1)}{RATE_{max,i}}$,
where $RATE_{max,i}:\textrm{intrinsic rate of }t_{i}$}}\tabularnewline
 &  & {\scriptsize{$t_{i}:\textrm{traffic proportion of applications,}$
($1\leq i\leq N$|$N:\textrm{number of applications}$)}}\tabularnewline
 &  & {\scriptsize{$\underset{i=1}{\overset{N}{\sum}}t_{i}=1$ and $\textrm{maximal rate requirement: }RATE_{max}=\underset{i=1}{\overset{N}{\sum}}t_{i}\cdot RATE_{max,i}$}}\tabularnewline
\hline 
\end{tabular}
\end{table*}

In this section, we numerically evaluate the performance of the proposed
optimal algorithm for the utility max-min fair link adaptation in
MATLAB. The downlink channel setup is modeled based on the TGn channel
models. We use profile B (residential) of those models in our simulations.
In order to ensure that the model is appropriate for IEEE 802.11ac
scenarios, we modified some of its parameters, including the Doppler
component, the angle of arrival (AoA), and the angle of departure
(AoD) according to the IEEE 802.11ac task group recommendations. The
system has been configured to operate at the 5.25-GHz carrier frequency
with a bandwidth of $W=20$ MHz subdivided into 64 subcarriers, 52
of which are used to carry data. There are nine different MCSs, which
results in transmission rates (i.e., $rate(m_{r}^{\theta})$) ranging
from 6.5 Mbps (BPSK, 1/2) to 78 Mbps (256-QAM, 5/6). The transmitter
is equipped with four antennae, and each of the four receivers has
only one antenna. We exploit four distinct utility functions, VoIP,
video streaming, file transfer, and online gaming applications, using
suggested functions in \cite{WH.Wang-app-alg-fair} and \cite{liu2007util-based-BW-Alloc}.
These utility functions are described in Table \ref{tab:Util-Functions}.
The calibration parameters for video streaming, file transfer and
online gaming are set according to \cite{liu2007util-based-BW-Alloc}.
The VoIP utility function is calibrated by the following parameters
(refer to \cite{headquarterscisco} for details): $L_{T}=3$, $\alpha_{i}\in\{\alpha_{1}=0.92,\alpha_{2}=0.95,\alpha_{3}=1\}$,
$l_{i}\in\{l_{1}=\{[21,32)\},l_{2}=\{[32,88)\},l_{3}=\{[88,\infty)\}\}$
(all in Kbps) and $U_{min,1}=0.7$ (for other receivers, $U_{min,2}=0.5$,
$U_{min,3}=0.4$ and $U_{min,4}=0.4$) . We have concentrated on the
long-time simulation scale of 20000 DL-MU transmissions, and for the
presented results in this section, the width of the 95\% confidence
interval of the true mean is less than 5\% of each plotted value.

The proposed scheme is compared with the equal power allocation (EPA)
and maximum utility schemes. The EPA scheme allocates the same (max-min
fair) power level to all receivers, while the maximum utility scheme
selects the link adaptation policies in the manner of maximizing the
total utility gain (performance gain). First, we validate the proposed
scheme in terms of the Jain's fairness index for the objective function
$O_{r}^{\Xi}$'s gap between the minimum receiver utility $U_{min,r}$
and the increased max-min fair utility determined by applying policy
$\rho_{r}^{sel}$ in order to obtain a quantified measure. Fig. \ref{fig:Jain's-Fairness-Index}
illustrates the measured results for two compared schemes. The horizontal
axis is time steps (A-MPDU transmissions) and the vertical axis shows
the corresponding fairness index. The index ranges from $\tfrac{1}{R}$
(worst case) to $1$ (best case) and is maximized when all receivers
gain a max-min fair proportion of the utilities. We deduce that the
proposed scheme guarantees a similar utility gap among the receivers
while EPA performs worse with regard to fairness. Second, we indicate
the efficiency of the proposed scheme by presenting the average utility
gain of the proposed and maximum utility schemes. Fig. \ref{fig:Total-average-of}
demonstrates the averaged utility gains for each compared scheme (the
horizontal axis is same as in Fig. \ref{fig:Jain's-Fairness-Index}
and the vertical axis is for utility score). Moreover, Table \ref{tab:Distinct-Utility-Averages}
represents the utility of each receiver in average (to interpret the
results, note that: $U_{min,1}=0.7$, $U_{min,2}=0.5$, $U_{min,3}=0.4$
and $U_{min,4}=0.4$). We observe that the maximum utility scheme
selects the policies to maximize the total utility, resulting in the
best utility gain. In contrast, the proposed scheme (max-min fair)
achieves the lower yet relatively close (see Fig. \ref{fig:Total-average-of}
and Table \ref{tab:Distinct-Utility-Averages}) utility gain by paying
the price of providing the max-min fairness and utility gain efficiency
combined. We conclude that the proposed scheme achieves 95\% of the
best possible total utility gain. As analyzed before, the simulation
results numerically confirm and validate the max-min fair and efficient
utility gain of the receivers in the proposed scheme.

\section{Conclusions}

In this letter, we proposed a novel model for utility max-min fair
link adaptation in the IEEE 802.11ac DL-MU. We achieve the max-min
fairness among the receivers using the simple, accurate, and practical
link adaptation model and corresponding solution algorithms, which
have been analyzed in this letter. Via simulation experiments, the
solution to the proposed model scheme is investigated and validated
by quantitative comparisons. 
\begin{table}[t]
\caption{Receiver Utilities in Average for Two Compared Schemes. \label{tab:Distinct-Utility-Averages}}

\hspace{0.03\textwidth}%
\begin{tabular}{|c|c|c|c|c|}
\hline 
Scenario\textbackslash{}Average Utility & $U_{1}$ & $U_{2}$ & $U_{3}$ & $U_{4}$\tabularnewline
\hline 
\hline 
Proposed Scheme & 0.9727 & 0.7706 & 0.6636 & 0.6371\tabularnewline
\hline 
Maximum Utility Scheme & 0.8772 & 0.7774 & 0.6636 & 0.8820\tabularnewline
\hline 
\end{tabular}
\end{table}

\appendices{}

\bibliographystyle{IEEEtran}


\end{document}